\documentstyle[12pt]{article}

\def\baselinestretch{1.2}
\catcode`\@=11
\def\section{\@startsection {section}{1}{\z@}{3.ex plus 1ex minus
 .2ex}{2.ex plus .2ex}{\raggedright\large\bf}}
\def\subsection{\@startsection{subsection}{2}{\z@}{2.75ex plus 1ex minus
 .2ex}{1.5ex plus .2ex}{\raggedright\bf}}
\def\appendix{{\newpage\section*{Appendices}}\let\appendix\section%
        {\setcounter{section}{0}
        \gdef\thesection{\Alph{section}}}\section}
\catcode`\@=12
\newskip\humongous \humongous=0pt plus 1000pt minus 1000pt
\def\caja{\mathsurround=0pt}
\def\eqalign#1{\,\vcenter{\openup1\jot \caja
        \ialign{\strut \hfil$\displaystyle{##}$&$
        \displaystyle{{}##}$\hfil\crcr#1\crcr}}\,}
\newif\ifdtup

\def\oldreffmt#1{\rlap{[#1]} \hbox to 2\parindent{}}

\def\figfmt#1{\rlap{Figure {#1}} \hbox to 1in{}}

\def\ltap{\raisebox{-.4ex}{\rlap{$\sim$}} \raisebox{.4ex}{$<$}}
\def\gtap{\raisebox{-.4ex}{\rlap{$\sim$}} \raisebox{.4ex}{$>$}}

\def\beq{\begin{equation}}
\def\eeq{\end{equation}}
\def\bea{\begin{eqnarray}}

\def\com#1#2{
        \left[#1, #2\right]}
\def\eea{\end{eqnarray}}
\def\ap#1,#2,#3#4{           {\it Ann. Phys. (NY)\/ }{\bf #1} (19#3#4) #2}
\def\apj#1,#2,#3#4{          {\it Astrophys. J.\/ }{\bf #1} (19#3#4) #2}
\def\apjl#1,#2,#3#4{         {\it Astrophys. J. Lett.\/ }{\bf #1} (19#3#4) #2}
\def\app#1,#2,#3#4{          {\it Acta Phys. Polon.\/ }{\bf #1} (19#3#4) #2}
\def\com#1,#2,#3#4{          {\it Comm. Math. Phys.\/ }{\bf #1} (19#3#4) #2}
\def\ib#1,#2,#3#4{           {\it ibid.\/ }{\bf #1} (19#3#4) #2}
\def\nat#1,#2,#3#4{          {\it Nature (London)\/ }{\bf #1} (19#3#4) #2}
\def\np#1,#2,#3#4{           {\it Nucl. Phys.\/ }{\bf B#1} (19#3#4) #2}
\def\npps#1,#2,#3#4{         {\it Nucl. Phys. B (Proc. Suppl.)\/ }{\bf B#1}
                             (19#3#4) #2}
\def\pl#1,#2,#3#4{           {\it Phys. Lett.\/ }{\bf #1B} (19#3#4) #2}
\def\pla#1,#2,#3#4{          {\it Phys. Lett.\/ }{\bf #1A} (19#3#4) #2}
\def\pr#1,#2,#3#4{           {\it Phys. Rev.\/ }{\bf D#1} (19#3#4) #2}
\def\prep#1,#2,#3#4{         {\it Phys. Rep.\/ }{\bf #1} (19#3#4) #2}
\def\prl#1,#2,#3#4{          {\it Phys. Rev. Lett.\/ }{\bf #1} (19#3#4) #2}
\def\pro#1,#2,#3#4{          {\it Prog. Theor. Phys.\/ }{\bf #1} (19#3#4) #2}
\def\rmp#1,#2,#3#4{          {\it Rev. Mod. Phys.\/ }{\bf #1} (19#3#4) #2}
\def\sp#1,#2,#3#4{           {\it Sov. Phys.-Usp.\/ }{\bf #1} (19#3#4) #2}

\def\zp#1,#2,#3#4{           {\it Zeit. fur Physik\/ }{\bf #1} (19#3#4) #2}
\catcode`\@=11
\def\eqnarray{\stepcounter{equation}\let\@currentlabel=\theequation
\global\@eqnswtrue
\global\@eqcnt\z@\tabskip\@centering\let\\=\@eqncr
\gdef\@@fix{}\def\eqno##1{\gdef\@@fix{##1}}%
$$\halign to \displaywidth\bgroup\@eqnsel\hskip\@centering
  $\displaystyle\tabskip\z@{##}$&\global\@eqcnt\@ne
  \hskip 2\arraycolsep \hfil${##}$\hfil
  &\global\@eqcnt\tw@ \hskip 2\arraycolsep $\displaystyle\tabskip\z@{##}$\hfil
   \tabskip\@centering&\llap{##}\tabskip\z@\cr}

\def\@@eqncr{\let\@tempa\relax
    \ifcase\@eqcnt \def\@tempa{& & &}\or \def\@tempa{& &}
      \else \def\@tempa{&}\fi
     \@tempa \if@eqnsw\@eqnnum\stepcounter{equation}\else\@@fix\gdef\@@fix{}\fi
     \global\@eqnswtrue\global\@eqcnt\z@\cr}

\newcount\hour
\newcount\minute
\newtoks\amorpm
\hour=\time\divide\hour by 60
\minute=\time{\multiply\hour by 60 \global\advance\minute by-\hour}
\edef\standardtime{{\ifnum\hour<12 \global\amorpm={am}%
	\else\global\amorpm={pm}\advance\hour by-12 \fi
	\ifnum\hour=0 \hour=12 \fi
	\number\hour:\ifnum\minute<10 0\fi\number\minute\the\amorpm}}
\edef\militarytime{\number\hour:\ifnum\minute<10 0\fi\number\minute}

\def\draftlabel#1{{\@bsphack\if@filesw {\let\thepage\relax
   \xdef\@gtempa{\write\@auxout{\string
      \newlabel{#1}{{\@currentlabel}{\thepage}}}}}\@gtempa
   \if@nobreak \ifvmode\nobreak\fi\fi\fi\@esphack}
        \gdef\@eqnlabel{#1}}
\def\@eqnlabel{}
\def\@vacuum{}
\def\marginnote#1{}
\def\draftmarginnote#1{\marginpar{\raggedright\scriptsize\tt#1}}
\def\draft{
	\pagestyle{plain}
	\overfullrule=2pt
        \oddsidemargin -.5truein
        \def\@oddhead{\sl \phantom{\today\quad\militarytime} \hfil
        \smash{\Large\sl DRAFT} \hfil \today\quad\militarytime}
        \let\@evenhead\@oddhead
        \let\label=\draftlabel
        \let\marginnote=\draftmarginnote
        \def\ps@empty{\let\@mkboth\@gobbletwo
        \def\@oddfoot{\hfil \smash{\Large\sl DRAFT} \hfil}
        \let\@evenfoot\@oddhead}
        \def\@eqnnum{(\theequation)\rlap{\kern\marginparsep\tt\@eqnlabel}%
        \global\let\@eqnlabel\@vacuum}  }

\def\theequation{\thesection.\arabic{equation}}
\@addtoreset{equation}{section}
\@addtoreset{footnote}{section}

\catcode`\@=12

\input epsf
\def\cpsbox{\epsfcheck\cpsbox}
\def\epsfcheck{\ifx\epsfbox\UnDeFiNeD
	\message{(NO epsf.tex, FIGURES WILL BE IGNORED)}
	\gdef\cpsbox##1##2{\vbox to 2in{\hbox to ##1 {\hss} \vss}}
\else\gdef\cpsbox##1##2{
	\setlength{\epsfxsize}{##1}
	\centerline{\epsfbox{##2}}}\fi}
\def\psinsert#1#2#3{
        \begin{figure}
  	\cpsbox{#1 \hsize}{#2}
	\medskip
	\centerline{
	   \vbox{\hsize=#1\hsize \footnotesize \def\baselinestretch{1.} #3 } }
        \end{figure}
}

\def\lae{\smash{\,\lower .5 ex \hbox{$\,\stackrel<\sim\,$}}}
\def\gae{\smash{\,\lower .5 ex \hbox{$\,\stackrel>\sim\,$}}}

\def\L{{\cal L}}

\def\beq{\begin{equation}}
\def\eeq{\end{equation}}

\def\sutw{${\rm SU}(2)_W$}

\def\KKb{$K^0-\bar K^0$}
\def\DDb{$D^0-\bar D^0$}

\begin{document}
\begin{titlepage}
\begin{center}
December 20, 1993\hfill    WIS--93/119/Dec--PH

\vskip 1 cm

{\large \bf  Bounds on Vector Leptoquarks}

\vskip 1 cm

Miriam Leurer

\vskip 1 cm

{\em Department of Particle Physics\\
The Weizmann Institute\\
Rehovot 76100\\
ISRAEL}

\end{center}

\vskip 1 cm

\begin{abstract}

We derive bounds on vector leptoquarks coupling to the first generation, using
data from low energy experiments as well as from high energy accelerators.
Similarly to the case of scalar leptoquarks, we find that the strongest
indirect bounds arise from atomic parity violation and universality in leptonic
$\pi$ decays. These bounds are considerably stronger than the first direct
bounds of HERA, restricting vector  leptoquarks that couple with
electromagnetic strength to right-handed quarks to lie above 430$~$GeV or
460$~$GeV,  and leptoquarks that couple with electromagnetic strength to
left-handed quarks to lie above 1.3$~$TeV, 1.2$~$TeV and 1.5$~$TeV for the
\sutw{} singlet, doublet and triplet respectively.

\end{abstract}
\end{titlepage}
\newpage
\section{Introduction}

The ongoing leptoquark search at the electron--proton machine HERA has
stimulated renewed interest in these particles and their phenomenology.
We have recently studied relevant data from low and high energy
experiments in order to deduce bounds on the couplings of {\it scalar}
leptoquarks \cite{me}. Here we shall do the same for {\it vector} leptoquarks.

As in the case of the scalars, we are interested in the {\it
unavoidable} bounds on the leptoquark couplings to the first generation. These
are the relevant couplings for HERA as well as for many other leptoquarks
searches. We find that the strongest bounds arise from low energy experiments:
Atomic parity violation, and universality in leptonic $\pi$ decays. Our bounds
are stronger than the first HERA results \cite{HERA} and they also have
important implications for various proposals for future indirect leptoquarks
searches in colliders \cite{methods}, as they already exclude significant
portions of the
region in parameter space that such searches can penetrate.

The paper is organized as follows: In the next section the vector leptoquarks
multiplets and their couplings are presented. In section 3 we review the bounds
from direct leptoquark searches and in section 4 we derive the indirect bounds
from atomic parity violation and universality in leptonic $\pi$ decays. Section
5 reviews bounds that turn out to be less useful than those of section 4.
Section 6 summarizes our results.

\section{The vector leptoquarks and their interactions}

The list of all possible vector leptoquark multiplets \cite{Buch2} includes the
$S$ and the $\tilde S$ leptoquarks in the $(0)_{-2/3}$ and $(0)_{-5/3}$
representations of $SU(2)_{\rm W}\times U(1)_{\rm Y}$, the $D$ and $\tilde D$
leptoquarks in the $(1/2)_{5/6}$ and $(1/2)_{-1/6}$ representations, and the
$T$ leptoquark in the $(1)_{-2/3}$ representation. Note that the scalar
leptoquark multiplets \cite{Buch2} also include two \sutw{} scalars, two
doublets and one triplet. The scalar and vector leptoquark multiplets differ
however in two important points: First, they carry different weak
hypercharges. Second, they carry different fermion numbers: $F=3B+L$ (with $B$
being baryon number and $L$ lepton number) vanishes for the \sutw{} doublet
scalar leptoquarks but is $(-2)$ for the \sutw{} doublet vectors.
The opposite happens for the \sutw{} singlets and triplet: here $F$ vanishes
for the vectors and $F=-2$ for the scalars.

As in the case of the scalar leptoquarks, we evade the strongest
bounds on the vector leptoquarks by demanding that they have no diquark
couplings, and that they couple chirally and diagonally to the first
generation.
We briefly repeat the discussion of the reasons for these demands:
\newline
$\bullet$ Diquark couplings are forbidden since these lead to nucleon decay
\cite{GG} and therefore imply that the leptoquark mass is of the order of the
GUT scale.
\newline
$\bullet$ Chirality of the couplings means that the leptoquark couples either
to left--handed (LH) quarks or to right--handed (RH) quarks, not to both. This
requirement is due to the observation \cite{Shanker} that a nonchiral
leptoquark that couples to the first generation gives a particularly enhanced
contribution to $\pi\longrightarrow e\nu$. To avoid a conflict with the
observed
universality in leptonic $\pi$ decays, the nonchiral vector leptoquark must
obey:
\beq
M/\sqrt{|g_Lg_R|} \geq 200~{\rm TeV},
\label{gLR}
\eeq
with $M$ the leptoquark mass and $g_L$, $g_R$ the couplings to the LH and RH
quarks respectively. This means that the leptoquark is very heavy or has  very
small couplings, and is consequently out of reach for present and near future
colliders. The bound of equation (\ref{gLR})  is four times stronger
than the analogous bounds for scalar leptoquarks \cite{me}. We shall see that
in general, vector leptoquark contributions to various processes
are enhanced relative to the scalar leptoquark contributions, although this
will not necessarily imply that the bounds
on the vector leptoquarks are stronger.

Some of the leptoquarks that are listed in the beginning of this section are
forced by their \sutw$\times U(1)_Y$ properties to be chiral. These are the
$\tilde S$ and the $\tilde D$ that can couple only to RH quarks, and the $T$
that can couple only to LH quarks. The other leptoquarks multiplets, the $S$
and the $D$, could couple both to LH and to RH quarks, but since we require
that couplings be chiral, we will from now on distinguish the $S_L$ and $D_L$
that couple to LH quarks from the $S_R$ and $D_R$ that couple to RH quarks.
\newline
$\bullet$ Diagonality of the leptoquark couplings means that the leptoquark
couples to a single generation of quarks and to a single generation of leptons.
For HERA we are interested in the case where the leptoquarks couple only to the
first generation. If this requirement is not fulfilled, the leptoquark induces
flavour changing neutral currents that lead to very strong bounds on its
parameters \cite{PS,Buch1}. In previous works \cite{oldme,me} we have pointed
out that strict diagonality is not really possible for leptoquarks that couple
to LH quarks, since the couplings to the down quarks are CKM rotated relative
to the couplings to the up quarks. It is however possible to demand that such
leptoquarks are approximately diagonal that is, they couple mainly to the first
generation, with their couplings to the second and third generations suppressed
by $O(\sin\theta_C)$ and $O(|V_{13}|+|V_{12}||V_{23}|)$ respectively, $V$ being
the CKM matrix.

The chirality and diagonality demands are very unlikely to be satisfied if the
vector leptoquarks are gauge bosons: to see this, note that leptoquarks carry
colour. If they are gauge bosons, the gauge symmetry must be some extension of
$SU(3)_C$, so that the leptoquarks together with the gluons are the gauge
bosons of the extended group. If one now requires that the leptoquarks couple
diagonally and chirally, these requirements must apply to the gluons as well;
namely, the gluons couple to the first generation only, and furthermore, to
quarks of a particular chirality only. This means that the theory should have
at least two sets of gluons -- those associated with the extended
gauge group of the first generation quarks of the particular chirality, and
those that are associated with all other quarks. There must then be some
mechanism to break the two colour groups to the diagonal one, leaving us with
the usual single set of massless gluons. We now face several problems: First,
each of the two colour groups is anomalous due to the chirality requirement,
and one needs to further extend the theory, adding fermions that will cancel
the anomalies. Second, one must also extend the standard model Higgs sector
in order to account for the masses of the first generation quarks and their
mixing with quarks of other generations. At this stage the model building task
becomes too tedious and the result too cumbersome to be convincing. With these
arguments in mind, we will in the following think of the vector
leptoquarks as composites rather than fundamental particles.

In addition to our requirements on the leptoquark couplings, we also
make some simplifying assumptions on the leptoquark spectrum: we assume that
there is at most one leptoquark multiplet, and that the mass splitting within
this multiplet is negligible. These assumptions simplify the presentation
of the results since they leave us with only two parameters: the leptoquark
multiplet mass, $M$, and its coupling to the first generation, $g$.

There is a significant difference between the requirements on the leptoquark
couplings and the assumptions on the leptoquark spectrum. If the requirements
on the leptoquark couplings are satisfied, the most severe bounds on the
leptoquark parameters are circumvented and we can concentrate on those bounds
which are absolutely unavoidable; if these requirements are not satisfied, the
bounds on the first generation couplings will just become stronger. In
contrast, the assumptions that the leptoquark spectrum is a single
multiplet, and that the mass splitting within the multiplet can be ignored, are
made for convenience. If these assumptions do not hold, the bounds can change
in
either direction -- they can become somewhat weaker or somewhat stronger, but
as discussed in \cite{me}, dramatic changes are unlikely.

We now introduce our notation:
the couplings of the leptoquarks that couple to RH quarks are given by
\begin{eqnarray}
\L_{S_R} &=& g~\bar e\,\gamma^\mu \, d_R  \, S_{\mu}^{(-2/3)}\nonumber\\
\L_{\tilde S} &=& g~\bar e\,\gamma^\mu\, u_R  \,\tilde
S_\mu^{(-5/3)}\nonumber\\
\L_{D_R} &=& g~\left(\bar \nu^c \,\gamma^\mu\, d_R D_\mu^{(1/3)} +
                \bar e^c\,\gamma^\mu\,  d_R D_\mu^{(4/3)} \right) \nonumber\\
\L_{\tilde D} &=& g~\left(\bar \nu^c \,\gamma^\mu\, u_R \tilde D_\mu^{(-2/3)} +
              \bar e^c\,\gamma^\mu\,  u_R\tilde D_\mu^{(1/3)} \right) \;,
\label{gR}
\end{eqnarray}
where the superscripts on the leptoquark fields indicate their electromagnetic
charge. In the case of the vector leptoquarks that couple to LH quarks
we have to introduce two sets of couplings: $g_i$ is the
coupling to the $i$'th up-quark generation, $g_i'$ is the coupling  to the
$i$'th down-quark generation, and they are related by the CKM rotation
$g_i'=g_j V_{ji}$.
\begin{eqnarray}
\L_{S_L} &=&
\sum_i \, \left(g_i~\bar \nu \,\gamma^\mu\,  u^i_L +
                g_i'~\bar e \,\gamma^\mu\,   d^i_L \right)
               \, S_\mu^{(-2/3)}\nonumber\\
\L_{D_L} &=&
\sum_i \, \left\{g_i~\bar e^c\,\gamma^\mu\,  u^i_L D_\mu^{(1/3)} +
                g_i'~\bar e^c\,\gamma^\mu\,  d^i_L D_\mu^{(4/3)} \right\}
\nonumber\\
\L_T &=& \sum_i \, \left\{
{\sqrt2} g_i~\bar e\,\gamma^\mu\,   u^i_L T_\mu^{(-5/3)}
+ (g_i~\bar \nu \,\gamma^\mu\,  u^i_L - g_i'~\bar e \,\gamma^\mu\,   d^i_L) \,
T_\mu^{(-2/3)}\right. \nonumber   \\
&&~~~~~~~~~~~~~~
\left. + {\sqrt2} g_i'~\bar \nu\,\gamma^\mu\,   d^i_L T_\mu^{(1/3)} \right\}
\; .
\label{YukawaL}
\end{eqnarray}
For these leptoquarks we define:
\beq
g=\sqrt{\sum_i~|g_i|^2}=\sqrt{\sum_i~|g'_i|^2}
\label{goverall}\;.
\eeq
$g$ is the overall strength of the Yukawa couplings, and our results are given
as bounds in the $g$ -- $M$ plane. Note that the first generation couplings are
equal to $g$ to a very good approximation (up to $2-3\%$), since we require
that the second and third generation couplings are suppressed by
$O(\sin\theta_C)$ and $O(|V_{13}|+|V_{12}\cdot V_{23}|)$. In the following the
differences between $g$, $g_1$ and $g_1'$ will be ignored.

We also introduce the parameters $\eta_I$, with $I$ running
over all leptoquark multiplets: $I=S_L,S_R,\tilde S,D_L,D_R,\tilde D,T$.
$\eta_I$ gets the value $1$ when we consider a theory with the leptoquark $I$,
and otherwise it vanishes.

\section{Direct bounds}

The LEP experiments have searched for {\it scalar} leptoquark pair production
in $Z$ decays. No evidence for such a decay mode was found and consequently LEP
set a lower bound on the scalar leptoquark mass: $M\gtap M_Z/2$
\cite{LEPdirect}. Since the signature of a vector leptoquark pair is very
similar to that of a pair of scalar leptoquarks, the LEP bound applies to
vector leptoquarks as well.

UA2 \cite{UA2} and CDF \cite{CDFdir} searched for first generation scalar
leptoquark pairs produced via an intermediate gluon. No events were seen, so
UA2 and CDF derived bounds on the leptoquark masses. The bounds depend on $b$,
the branching ratio of the leptoquark decay to $e^\pm$ and a jet, since the
hadronic colliders experiments cannot identify events in which both leptoquarks
decayed to a neutrino and a jet, and CDF also cannot identify an event in which
one of the leptoquarks decayed to a neutrino and a jet. The CDF bounds on
scalar leptoquarks have been recently translated to bounds on vector
leptoquarks \cite{HRP}. The bounds on the vectors depend not only on $b$, but
also on the ``anomalous chromomagnetic moment'' of the leptoquarks which
affects
significantly the leptoquark production cross section. Here we will use only
the
weakest bounds that apply in the case of vanishing anomalous chromomagnetic
moment: $M\gtap 150~$GeV for $b=1/2$ and $M\gtap 180~$GeV for $b=1$. The $S_L$
vector leptoquark has $b=1/2$ and therefore only the weaker bound
$M\gtap150~$GeV applies to it. All the other vector leptoquark multiplets
contain at least one component with $b=1$. Using our assumption of no mass
splitting within a multiplet we therefore find that all the vector leptoquarks,
but $S_L$, are heavier than 180$~$GeV.

\section{Indirect bounds}

In this section we will discuss the strongest indirect bounds that we find for
vector leptoquarks. These arise from two low energy experiments: Atomic parity
violation and universality in leptonic $\pi$ decays.

Atomic parity violation in Cesium is experimentally measured and theoretically
calculated to a high accuracy. It has been advocated for some time that this
process should give strong bounds on leptoquarks \cite{Lang}, and in
\cite{me} we found that this was indeed the case for scalar leptoquarks. We now
repeat the analysis for the vectors. We look at the
Cesium ``weak charge'' defined by:
\beq
Q_W=-2\left[C_{1u}(2Z+N) + C_{1d}(2N+Z)\right]
\label{QW}\;,
\eeq
with $C_{1u}$ and $C_{1d}$
defined {\it e.g.} in \cite{PDG} and with $Z=55$ and $N\simeq78$ for Cesium.
The latest experimental result \cite{Csexp} and the standard model estimate
\cite{Cstheo} for $Q_W$ are:
\begin{eqnarray}
Q_W^{\rm exp}&=&-71.04\pm1.81 \nonumber\\
Q_W^{\rm SM}&=&-73.12\pm0.09
\label{qwexpsm}\;.
\end{eqnarray}
In a theory with a vector leptoquark, there is an additional contribution to
$Q_W$, given by:
\beq
\eqalign{
\Delta Q_W^{LQ}=4\left(\frac {g/M}{g_W/M_W}\right)^2
\left[\vphantom{(Z+2N)\eta_{S_L}}\right.&
(2Z+N) \cdot(\eta_{\tilde S}-\eta_{D_L}+\eta_{\tilde D}-2\eta_T)  \cr
             &\left. +
(Z+2N) \cdot(-\eta_{S_L}+\eta_{S_R}-\eta_{D_L}+\eta_{D_R}-\eta_T)
\right]
\label{qwlq}
}
\eeq
Here $g$ and $M$ are the coupling and mass of the leptoquarks and $g_W$ and
$M_W$ are the coupling and mass of the $W$ boson.
The close agreement between the experimental $Q_W$ value and the standard model
estimate (see equation (\ref{qwexpsm})) leads to strong bounds on $g/M$.
These are summarized in table 1.

The vector leptoquark contribution to $Q_W$ can be derived from the
scalar leptoquark contribution of \cite{me} by: (i) Exchanging $Z$ and $N$;
(ii) multiplying by a $(-)$ sign and (iii) enhancing the contribution by a
factor of 2. Despite of this enhancement, atomic parity violation bounds on
vector
leptoquarks are not always stronger than the corresponding bounds on the scalar
leptoquarks. This is due to the sign of the leptoquark contribution, which has
a significant effect on the bound.

\begin{table}
\begin{center}
\begin{tabular}{|l|c|c|c|c|c|c|c|}\hline
&$S_L$&$S_R$&$\tilde S$&$D_L$&$D_R$&$\tilde D$&$T$\\ \hline
$M_{4\pi}$&
10000 & 5300 & 5000 & 14000 & 5300 & 5000 & 17000\\ \hline
$M_1$&
2900 & 1500 & 1400 & 4100 & 1500 & 1400 & 4900\\ \hline
$M_e$&
890  & 460  & 430  & 1200 & 460  & 430  & 1500\\ \hline
\end{tabular}
\medskip
\caption[table1]{
\it Atomic parity violation 95\% CL lower bounds on the ratio $M/g$,
in GeV. The bounds  are presented in three equivalent ways:
$M_{4\pi}$ is the lower bound on the leptoquark mass when the coupling becomes
nonperturbative $g^2=4\pi$, $M_1$ is the bound when the coupling is 1, and it
is thus the bound on $M/g$, and $M_e$ is the bound when the coupling is equal
to
the electromagnetic coupling $g=e$.}
\end{center}
\end{table}

Universality in leptonic $\pi$ decays had been used to derive a bound on the
scalar leptoquark $S_L$ already in 1986 \cite{Buch1}. In \cite{me} we updated
this bound and added the corresponding bound for the $T$ scalar leptoquark.
Here we repeat the analysis and find bounds on the $S_L$ and $T$ {\it vector}
leptoquarks. The quantity that is measured and calculated is
$R=BR(\pi\longrightarrow e\nu)/BR(\pi\longrightarrow \mu\nu)$.
There are two recent measurements of $R$, one by TRIUMF\cite{TRIUMPH}, the
other by PSI \cite{PSI}. Combining their results we
find:
\beq
R^{\rm exp}=(1.2310\pm0.0037)\cdot 10^{-4}\;.
\label{Rexp}
\eeq
The theoretical standard model calculation by Marciano and Sirlin has been
updated \cite{MaSi} and the error is considerably reduced:
\beq
R^{SM}=(1.2352\pm0.0005)\cdot 10^{-4}\;.
\label{Rthe}
\eeq
The theoretical prediction in a theory with a vector leptoquark is:
\beq
R^{LQ}=R^{SM}
\left(1+2\left(\frac {g/M}{g_W/M_W}\right)^2\cdot(\eta_{S_L}-\eta_T)\right)^2
\label{Rlq}
\eeq
Equations (\ref{Rexp}--\ref{Rlq}) lead to the bounds of table 2. Note that
leptonic $\pi$ decays provide the strongest bound on the $S_L$ vector
leptoquark, while atomic parity violation supplies the strongest bounds for all
other vector leptoquarks, including the $T$. Note also that, again, the vector
leptoquark contribution is enhanced by a factor of 2 relative to that of the
scalar leptoquarks.
\begin{table}
\begin{center}
\begin{tabular}{|l|c|c|}\hline
&$S_L$&$T$\\ \hline
$M_{4\pi}$&
15500 & 9000\\ \hline
$M_1$&
4400 & 2500\\ \hline
$M_e$&
1300 & 760\\ \hline
\end{tabular}
\medskip
\caption[table2]{\it 95\% CL bounds on the ratio $M/g$, in GeV, from
universality in
leptonic $\pi$ decays.}
\end{center}
\end{table}

It is interesting to observe that the two bounds discussed in this section
reflect the consequences of our assumptions -- the chirality and diagonality of
the leptoquark couplings: Chirality of the leptoquark couplings implies that
processes mediated by these particles violate parity, while  diagonality of the
couplings implies that the leptoquarks distinguish the generations and
may therefore induce deviations from universality.

\section{Other bounds}
In this section we will discuss various processes that give weaker bounds on
vector leptoquarks than those of atomic parity violation and leptonic $\pi$
decays.

{\bf FCNC processes:} In \cite{oldme,me} we showed that FCNC processes
can give a significant bound on scalar leptoquarks that couple to LH quarks.
This was based on three main observations: The first observation is that FCNC
processes are unavoidable for leptoquarks that couple to LH quarks. The second
observation is that if one has FCNC bounds from both quark sectors it is
possible to combine them to a bound on the overall coupling $g$. The last
observation, which is troublesome in the case of the vector leptoquarks, is
that there are indeed FCNC bounds from both sectors: It is well known that
there are FCNC bounds in the down sector, which arise from rare $K$ decays. The
fact that there is also a significant FCNC bound in the up sector was
pointed out in \cite{oldme}, where the one loop contributions of leptoquarks to
\DDb{} and \KKb{} mixing were discussed. The problem with vector leptoquarks
is that the one-loop calculation is not a clear procedure: we have pointed out
that a vector leptoquark is unlikely to be fundamental. If it is
composite, its loop contribution to neutral meson mixing diverges
and it should be cutoff at the compositeness scale. This  cutoff procedure is
not well defined since we do not know what is the appropriate compositeness
scale to be used, although we believe it is similar in size to the leptoquark
mass; also, one should take into account other contributions that may arise
from the underlying theory, but are unknown to us. We therefore do not attempt
to extract bounds on vector leptoquarks from \DDb{} mixing, and have no bound
on $g$ from FCNC processes.

{\bf Bounds from other processes:} We have studied bounds that can arise from
$eD$ scattering, from the observed $e^+e^-$ mass distribution in
$p\bar p\longrightarrow e^+e^- + any$ and  from the hadronic forward backward
asymmetry in $e^+e^-$ machine. We find
that the case of vector leptoquarks is similar to that of scalar leptoquarks,
in that all these processes give weaker bounds than atomic parity
violation and leptonic $\pi$ decays. We now briefly review our results on these
processes.

$eD$ scattering probes the parity violating quantity
$C_{2u}-C_{2d}/2$. The contribution of a vector leptoquarks to this quantity
is given by:
\beq
\Delta(C_{2u}-C_{2d}/2)^{\rm LQ} = \left(\frac {g/M}{g_W/M_W}\right)^2
(-\eta_{S_L}+\eta_{S_R}-2\eta_{\tilde
S} - \eta_{D_L}-\eta_{D_R}+\eta_{\tilde D}
+2\eta_T)\;.
\label{eDlq}
\eeq
Comparing the experimental value ($-0.03\pm0.13$) to the standard model value
($-0.047\pm 0.005$) \cite{PDG}
leads to the bounds of table 3, which are considerably weaker than those of the
previous section.
\begin{table}
\begin{center}
\begin{tabular}
{|l|c|c|c|c|c|c|c|}\hline
&$S_L$&$S_R$&$\tilde S$&$D_L$&$D_R$&$\tilde D$&$T$\\ \hline
$M_{4\pi}$&
890 & 840 & 1270 & 890 & 890 & 840 & $1170$ \\ \hline
$M_1$&
250 & 240 & 360 & 250 & 250 & 240 & $330$ \\ \hline
$M_e$&
80 & 70 & 110 & 80 & 80 & 70 & $100$ \\ \hline
\end{tabular}
\medskip
\caption[table3]{\it $eD$ scattering 95\% CL bounds on $M/g$, in GeV.}
\end{center}
\end{table}

Turning to $p\bar p$ scattering to $e^+e^-$, we note that CDF \cite{CDF}
derived bounds of the order of 2$~$TeV on the compositeness scale by studying
the mass distribution of the electron--positron system. In \cite{me} we deduced
that similar bounds should apply to scalar leptoquarks, namely
$M_{4\pi}\gtap2~$TeV. Here we extend this to vector leptoquarks: The bound for
vector leptoquarks is stronger by a factor of $\sqrt2$ since the
coefficient of the four-Fermi operator, {\it e.g.} $\bar q_L\gamma^\mu q_L
\bar e_L\gamma^\mu e_L$, is enhanced by a factor of 2 relative to the case of
the scalars. The bound for vector leptoquarks therefore reads
$M_{4\pi}\gtap 3~$TeV. Note that for leptoquarks that couple to RH quarks
this bound is weaker only by a factor of $\sim2$ than the atomic parity
violation bound. It may therefore be worthwhile to repeat the CDF analysis
with more data and apply it specifically to vector leptoquarks.

\begin{table}
\begin{center}
\begin{tabular}
{|l|c|c|c|c|c|c|c|}\hline
&$S_L$&$S_R$&$\tilde S$&$D_L$&$D_R$&$\tilde D$&$T$\\ \hline
$M_{4\pi}$&
1300 & 800 & 1550 & 2700 & 1400 & 1950 & 1200\\ \hline
$M_1$&
380 & 230 & 440 & 750 & 400 & 550 & 340 \\ \hline
$M_e$&
110 & 70 & 130 & 230 & 120 & 170 & 100\\ \hline
\end{tabular}
\medskip
\caption[table4]{\it The 95\% CL lower bounds on $M/g$, in GeV,
derived from TRISTAN data. For the $\tilde D$ leptoquark there is
also a small allowed region at $142~{\rm GeV}\ltap M_1 \ltap 149~{\rm GeV}$}
\end{center}
\end{table}
Hadronic forward backward asymmetries in $e^+e^-$ machines: The process we look
at is $e^+e^-\longrightarrow q\bar q$, where a particular scattering is called
``forward'' if the negatively charged quark or antiquark scatters into the
forward hemisphere of the electron beam. Hadronic forward--backward asymmetry
was studied at PEP \cite{PEP}, in PETRA \cite{JADE}, in TRISTAN \cite{TRISTAN}
and in LEP \cite{LEP}. We concentrated on the results of TRISTAN and LEP, and
found that TRISTAN data gives the stricter bounds on the leptoquarks
parameters. Using the detailed data on differential cross sections provided to
us by TOPAZ and AMY,  we derived bounds on vector leptoquarks parameters by
comparing the experimentally measured differential cross section to the
prediction of the leptoquark theory. Our results are summarized in table 4. We
should note that the bounds in this table apply to heavy leptoquarks (of $\sim
1~$TeV and up). The bounds on the couplings of lighter leptoquarks are somewhat
weaker (by up to 6\%).
These bounds are again considerably weaker than the atomic parity violation
and the leptonic $\pi$ decay
bounds. We still find them interesting since they apply to any leptoquark that
couples chirally to the electron and to the first and/or the  second quark
generations. For the $S_R$ and the $D_R$ leptoquarks, these bounds apply
also when they couple to the $b$ quark of the third generation.

\section{Summary}
Our bounds on vector leptoquarks are summarized in table 5, which combines the
results of tables 1 and 2. Note in particular the last row in this table:
Vector
leptoquarks that couple with electromagnetic strength are excluded far above
HERA's kinematical limit of $300~$GeV (the weakest bound, applying to $\tilde
S$ and $\tilde D$, reads $M_e\geq 430~$GeV).

\begin{table}
\begin{center}
\begin{tabular}{|l|c|c|c|c|c|c|c|}\hline
&$S_L$&$S_R$&$\tilde S$&$D_L$&$D_R$&$\tilde D$&$T$\\ \hline
$M_{4\pi}$&
15500 & 5300 & 5000 & 14400 & 5300 & 5000 & 17400\\ \hline
$M_1$&
4400 & 1500 & 1400 & 4100 & 1500 & 1400 & 4900\\ \hline
$M_e$&
1300
& 460  & 430  & 1200 & 460  & 430  & 1500\\ \hline
\end{tabular}
\medskip
\caption[table5]{
\it Summary of the 95\% CL lower bounds on the ratio $M/g$,
in GeV, for vector leptoquarks.}
\end{center}
\end{table}

In figure 1 we compare our bounds with the first results from HERA in the mass
range that is bounded from below by the CDF direct bound ($150~$GeV for $S_L$
and $180~$GeV for all other leptoquark multiplets) and from above by HERA's
kinematical limit ($M\ltap 300~$GeV). Clearly, our bounds at the moment are far
more strict. In the future HERA's results should improve considerably and will
then win over our bounds in part of this mass range. As for higher leptoquark
masses, there are some suggestions in the literature to search for them in HERA
via indirect effects\cite{methods}. However, significant portions of the
regions in parameter space that can be penetrated into via indirect methods at
HERA (and at other colliders) are already excluded by our indirect low
energy bounds of table~5.

\psinsert{.9}{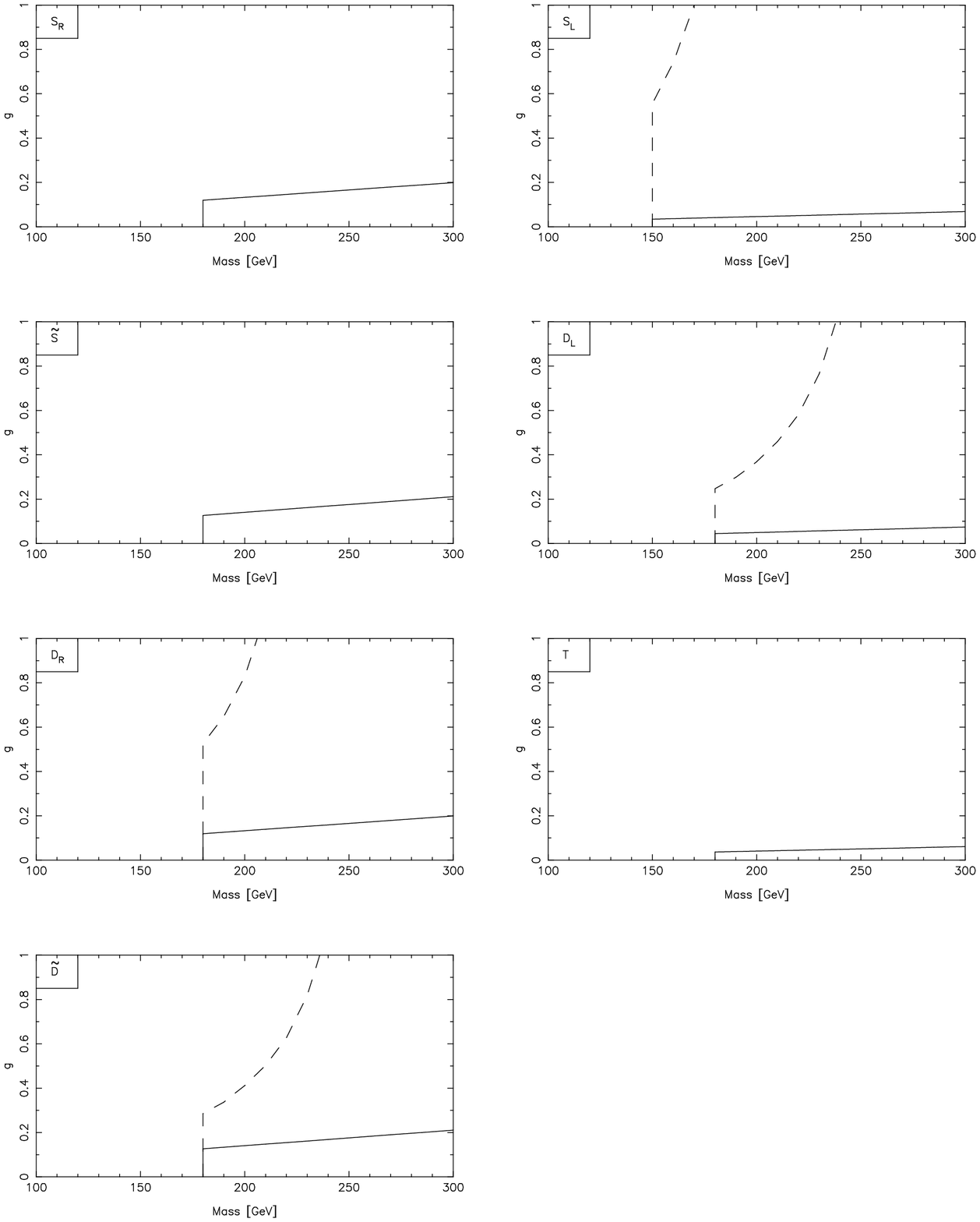}{{\bf Figure 1.} Our indirect bounds (full line)
compared with the direct bounds (dashed line) of the H1 group of HERA
\cite{HERA}. Note that for three of the leptoquark multiplet, $S_R$, $\tilde S$
and $T$, HERA does not yet provide any bounds in the mass region allowed by the
CDF direct bound $M\geq 180~$GeV.}

Finally, we wish to stress again that the bounds in table 5 are the weakest
possible bounds on vector leptoquarks, and apply to leptoquarks that couple
chirally and diagonally to the first generation. As we discussed in section 2,
{\it fundamental} vector leptoquarks (gauge bosons) are not likely to obey the
chirality and diagonality requirements. The bounds on their couplings are
therefore so strong that such particles are beyond the discovery limit
of present and near future colliders.

{\noindent
{\bf Acknowledgements:}
I am grateful to Neil Marcus and Ilan Levine for helpful discussions and to
Kazuo Abe and Ichiro Adachi who provided me with the data of AMY and TOPAZ.}

\end{document}